\documentclass[
aps,prd,
preprintnumbers,
amsmath,
amssymb,
nofootinbib,
preprintnumbers
]{revtex4}
\usepackage{here}
\usepackage{footnote}
\usepackage{comment,braket}
\usepackage{epsf}
\usepackage{amsmath}
\usepackage{graphics}
\usepackage{amsfonts}
\usepackage{amssymb}
\usepackage{latexsym}
\usepackage{color}
\usepackage{natbib}
\usepackage{graphicx}
\usepackage{hyperref}
\usepackage{array}

\usepackage[svgnames]{xcolor}
\definecolor{phthaloblue}{rgb}{0.0, 0.06, 0.54}
\hypersetup{
    colorlinks=true,
    linkcolor=blue,
    citecolor=blue,
    filecolor=blue,
    urlcolor=blue,
    }

\begin{document}
\title{From Entropy to Echoes:\\ Counting the quasi-normal modes and the quantum limit of silence}
\author{Naritaka Oshita$^1$ and Niayesh Afshordi$^{2,3,4}$}
\affiliation{$^1$RIKEN iTHEMS, Wako, Saitama, 351-0198, Japan}
\affiliation{$^2$Department of Physics and Astronomy, University of Waterloo, 200 University Ave W, N2L 3G1, Waterloo, Canada}
\affiliation{$^3$Waterloo Centre for Astrophysics, University of Waterloo, Waterloo, ON, N2L 3G1, Canada}
\affiliation{$^4$Perimeter Institute For Theoretical Physics, 31 Caroline St N, Waterloo, Canada}
\preprint{RIKEN-iTHEMS-Report-23}

\begin{abstract}
We estimate the canonical entropy of a quantum black hole by counting its quasi-normal modes.
We first show that the partition function of a classical black hole, evaluated by counting the quasi-normal modes with a thermodyanmic Boltzmann weight, leads to a small entropy of order unity due to the small contribution from higher angular modes. 
We then discuss how this will be modified when taking into account dissipation effects near the horizon  due to interaction with the quantum black hole microstates. The structure of quasi-normal modes drastically changes, yielding a fundamental frequency of the inverse of $t_{\rm echo} \sim$ log(Entropy)/Temperature. This is the time-scale for reflection from the microstates (or the quantum time limit of silence, followed by echoes), {\it independent of the strength of dissipation}, and is comparable to the scrambling time proposed by Sekino \& Susskind. Setting the dissipation constant to Planck time, we reproduce the Bekenstein-Hawking entropy of $\sim$ (Horizon area)/(Planck area). This result suggests the possibility of simulating black hole entropy in analog horizons realized in condensed matter systems.
\end{abstract}

\maketitle

\section{Introduction}
\begin{figure}[h]
\centering
\includegraphics[width=0.6\linewidth]{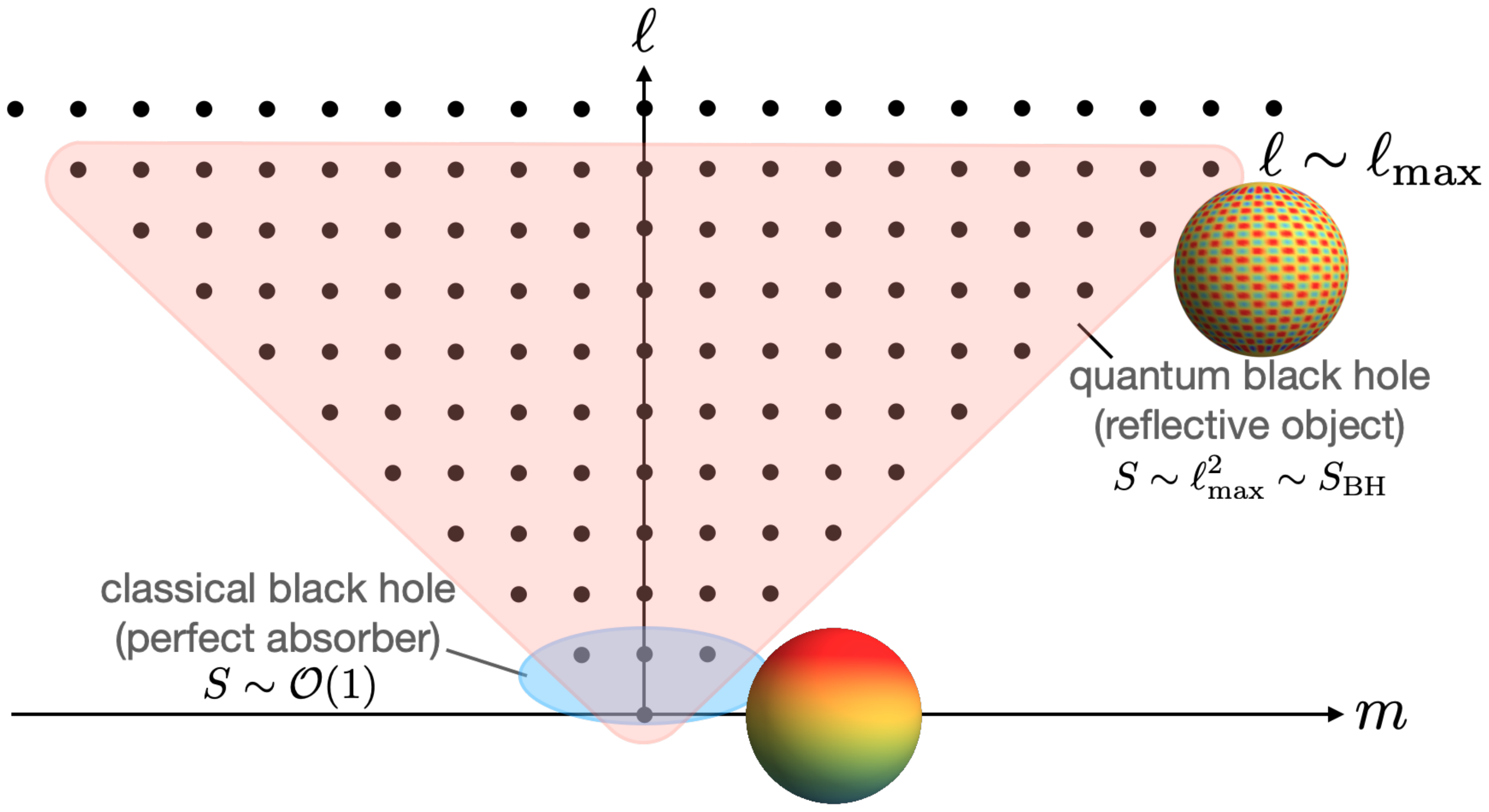}
\caption{Blue and red shaded angular modes significantly contribute to a classical and quantum black hole entropy, respectively. For a classical black hole, only lower angular modes (blue shaded) contribute to the entropy. On the other hand, for a quantum black hole, having a reflective boundary due to the dissipation near the horizon, not only the lower angular modes but also many higher modes contribute to it (red shaded).}
\label{fig:mode_count}
\end{figure}

A black hole is one of the simplest objects in the Universe, at least at a {\it macroscopic level}, as it is characterized by only three parameters, i.e., its mass, angular momentum, and electric charge. This is guaranteed by the no-hair theorem in general relativity. On the other hand, it is believed that, at a {\it microscopic level}, black holes have many quantum degrees of freedom, which are accounted by their large entropy. The strong gravity of a black hole leads to the formation of an event horizon that hides all interior information. The lost information from the exterior system may be associated with the Bekenstein-Hawking entropy of the black hole: $S= \frac{\cal A}{4G}$, where ${\cal A}$ is the event horizon area, $G$ is the Newton's constant, and we here take the natural unit $\hbar = c= 1$. However, we still do not know exactly what kind of degrees of freedom on the horizon leads to the Bekenstein-Hawking entropy. Although there are many candidates for the degrees of freedom \cite{Bekenstein:1973ur,Strominger:1996sh,Hod:1998vk,Immirzi:1996dr,Ashtekar:2000eq,Dreyer:2002vy,Meissner:2004ju,Dittrich:2005sy,Cadoni:2021jer}, we here consider the quasi-normal (QN) modes of a black hole as a possible origin of the entropy. That is, we assume that the quantum fluctuations of a black hole geometry can be represented by the superposition of the eigenstates associated with the QN modes in the analogy of a quantized harmonic oscillator. The QN frequencies, $\omega = \omega_{\ell mn} = \omega_{R,\ell mn} - i \omega_{I,\ell mn}$, are quite sensitive to the structure of spacetime in the vicinity of the horizon. The real part of a QN frequency $\omega_{R,\ell mn}$ is the frequency and the imaginary part $\omega_{I,\ell mn}$ is the damping rate of the QN mode. Even a small disturbance near the horizon may significantly affect the QN frequencies (e.g., \cite{Cardoso:2016rao,Jaramillo:2020tuu}).
Indeed, it has been thought that quantum effects at the horizon may cause a non-zero reflectivity and lead to the emission of gravitational wave (GW) echo after the ringdown emission \cite{Cardoso:2016rao,Abedi:2016hgu,Holdom:2016nek,Oshita:2018fqu,Oshita:2019sat}. The QN modes of the echoing black hole are long-lived due to the resonance near the horizon and are very different from the QN modes of a classical black hole.
In this paper, we evaluate the entropy of a non-spinning black hole by counting the QN modes of a classical and quantum black hole with the weight of the Boltzmann factor, i.e., $e^{-\beta_{\rm H}\tilde{\omega}_{\ell mn}}$, where $\beta_{\rm H}$ is the inverse of the Hawking temperature and $\tilde{\omega}_{\ell mn}$ is the proper frequency of $\tilde{\omega}_{\ell mn} \equiv |\omega_{\ell mn}|$ (see Ref. \cite{Maggiore:2007nq}). In other words, we here assume that QN modes may be thermally excited by the Hawking plasma \cite{Hawking:1974rv} surrounding the horizon. We find that for a classical black hole, only the lower $\ell$-modes contribute to the entropy, which is well below the Bekenstein-Hawking formula\footnote{In Ref. \cite{Cadoni:2021jer}, the authors propose the derivation of the Bekenstein-hawking formula by counting QN modes of a classical black hole. In the derivation, they assume that $N$ independent harmonic oscillators are assigned to each QN mode, but we do not assume it. This is a crucial difference between their derivation and ours.}. On the other hand,  the Bekenstein-Hawking entropy can be recovered from our entropy estimation with a quantum black hole, if we include dissipation near the horizon. In this case, the higher-$\ell$ modes significantly contribute to the entropy (see Figure \ref{fig:mode_count}).

This paper is organized as follow. In Sec. \ref{sec_counting_QN}, we discuss how we can estimate black hole entropy from the counting of QN modes. We first consider the classical black hole to show that the estimated entropy is of the order of unity. On the other hand, we find that for the quantum black hole, involving a reflective boundary near the horizon, the estimated entropy is of the order of the Bekenstein-Hawking entropy. In Sec. \ref{sec_discussions}, we provide our conclusion.

\section{counting of quasi-normal modes}
\label{sec_counting_QN}
We first show that the entropy estimation based on the counting of QN modes with the weight of the Boltzmann factor does not lead to the area law of black hole entropy\footnote{In ref. \cite{Cadoni:2021jer}, the authors discuss the derivation of the Bekenstein-Hawking entropy based on the entropy formula of multiple harmonic oscillators. It is motivated by the fuzzball proposal \cite{Lunin:2001jy,Mathur:2005zp}.} as higher harmonic modes have higher frequencies in the real part of their QN frequencies and is suppressed by the Boltzmann factors. In the latter part of this section, on the other hand, we consider a quantum black hole, whose QN modes are totally different from the original ones because of the dissipation effect of the horizon. We then show that in this case, higher harmonics modes contribute to the partition function, and the entropy turns out to be of the order of the Bekenstein-Hawking entropy.

\subsection{classical black holes}
For classical black holes, QN frequencies are given by the complex frequencies satisfying ingoing and outgoing boundary conditions at horizon and at infinity, respectively. A Schwarzschild black hole has infinite number of QN modes, that is, there are infinite number of overtones, labeled by a non-negative integer $n$, for each (spherical or spheroidal) harmonic mode. The harmonics modes are labeled by $(\ell, m)$ and infinitely exist with the restriction of $-\ell \leq m \leq \ell$. For a Schwarzschild black hole with higher harmonics ($\ell \gg 1$), the WKB approximation works well, and the approximated formula of QN mode reduces to \cite{Press:1971wr,Iyer:1986nq,Berti:2009kk}
\begin{align}
\begin{split}
\omega_{\ell mn} &= \omega_{R, \ell mn} -i \omega_{I, \ell mn},\\
\text{with} \ &\omega_{R, \ell mn} \simeq \Omega \left( \ell+\frac{1}{2} \right),\\
&\omega_{I, \ell mn} \simeq \Omega \left( n+\frac{1}{2} \right),
\end{split}
\end{align}
where $\Omega \equiv (3 \sqrt{3} G M)^{-1}$. Here we estimate the entropy of a static black hole by assuming it is in thermal equilibrium, whose temperature is $T_{\rm H} \sim (8 \pi GM)^{-1}$. A (thermodynamic) energy associated with each QN mode can be identified with the proper frequency $\tilde{\omega}_{\ell mn} \equiv \sqrt{\omega_{R, \ell mn}^2 + \omega_{I, \ell mn}^2}$ \cite{Maggiore:2007nq,Cadoni:2021jer}. Then, to compute the partition function of excited QN modes in a quasi-thermal Hawking plasma \cite{Hawking:1974rv}, we use the Boltzmann factor of $e^{-\beta_{\rm H} \tilde{\omega}_{\ell mn}}$, where $\beta_{\rm H} \equiv 1/T_{\rm H}$ is the inverse temperature. It is found that the partition function is given by
\begin{equation}
Z \sim \sum_{\ell mn} \exp\left[-\beta \Omega \sqrt{(\ell+1/2)^2 + (n+1/2)^2} \right] = \sum_{\ell mn} \exp\left[-\frac{8 \pi \beta}{3 \sqrt{3} \beta_{\rm H}} \sqrt{(\ell+1/2)^2 + (n+1/2)^2} \right].
\end{equation}
The canonical entropy of the excited modes of a static black hole reads
\begin{align}
S &= \ln Z -\beta \partial_{\beta} \ln Z|_{\beta = \beta_{\rm H}} \sim 0.32,
\end{align}
which is of order unity, $S \sim {\cal O} (1) \ll {\cal A}/4G$, and thus does not follow the area law. This is explained by the fact that only a few modes mainly contribute to the partition function; none of higher harmonics and higher overtones play a big role due to the exponential suppression of the Boltzmann factors.

\subsection{quantum black holes}
In this subsection, we will introduce a dissipation term in the propagation of perturbations around a black hole, as a phenomenological model for interaction with quantum microstructure near the horizon. This is motivated by the membrane paradigm \cite{Thorne:1986iy} where the horizon is replaced by the viscous fluid \footnote{It is sometimes referred as the stretched horizon \cite{Susskind:1993if}.}. Such a dissipative effect changes the dispersion relation of propagating modes near the horizon radius, which could also result in partial reflection of the modes \cite{Oshita:2018fqu,Oshita:2019sat}. This changes the boundary condition we impose to derive QN modes and the QN frequencies are drastically modified from the well known values in the literature \cite{Wang:2019rcf}. We will first briefly review the QN modes with a reflective boundary condition and show that even higher harmonics contribute to the partition function of a linearly-excited quantum black hole. We finally perform the estimation of the black hole entropy to show that it follows the area law unlike the classical situation, above.
\begin{figure}[h]
\centering
\includegraphics[width=0.5\linewidth]{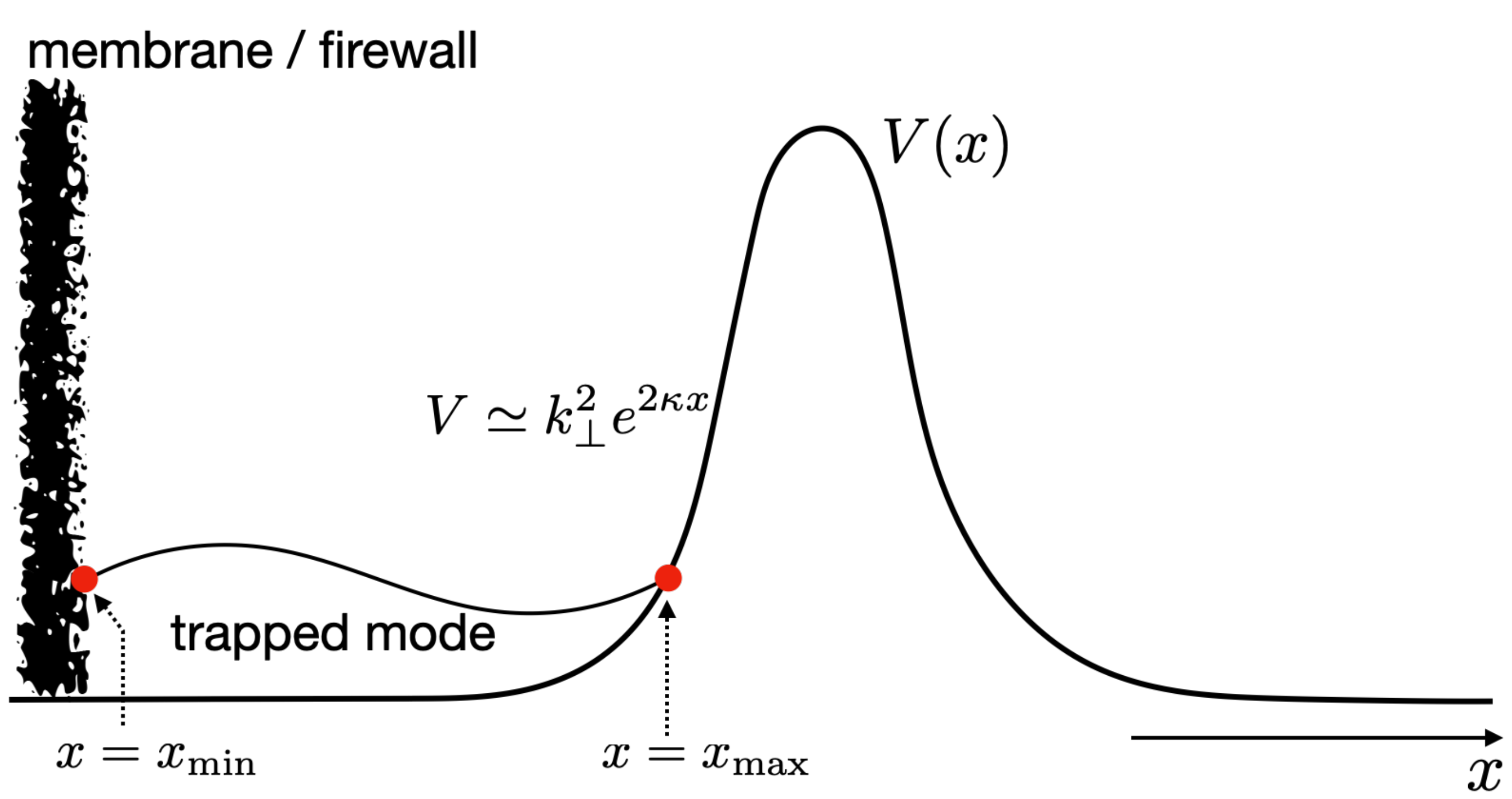}
\caption{Angular momentum barrier and a trapped mode in the vicinity of the black hole.}
\label{fig:potential}
\end{figure}

The reflection of propagating modes near the horizon radius in general forms a {\it cavity} between the reflective surface and angular momentum barrier (see Figure \ref{fig:potential}). Therefore, the QN frequencies are simply obtained by $\simeq \pi n/L_{\ell m}$, where $L_{\ell m}$ is the length of the cavity for the angular mode of $(\ell, m)$, and the overtone number $n$ is a positive integer. We will estimate the length of cavity, $L_{\ell m}$, in the following analysis. For simplicity, we start with the Rindler radar coordinates that gives the near-horizon geometry of a 3+1 dimensional black hole:
\begin{equation}
    ds^2= -\exp(2\kappa x) (-dt^2+dx^2) +dy^2+dz^2,
\end{equation}
where $x$ is the analogous to the tortoise radial coordinate, while $y$ and $z$ are angular directions parallel to the horizon, and $\kappa=2\pi T_{\rm H}$ is surface gravity, proportional to the horizon temperature $T_{\rm H}$. In the following, we will show that the dissipation modeling the viscosity of a membrane or stretched horizon modifies the dispersion relation at the near-horizon region, which leads to a partial reflection of propagating modes near horizon. Such a reflective boundary and the angular momentum barrier located outside of a black hole forms a cavity to trap low-frequency modes. It drastically modifies the structure of QN modes of a black hole and multiple normal modes are excited in the vicinity of the horizon. Then higher harmonics are excited up to $\ell \lesssim 2M/l_{\rm Pl}$ to contribute to the partition function, which gives the black hole entropy following the area law.

To simplify the following computation, let us consider a massless scalar field which satisfies the Klein-Gordon equation $\Box \phi =0$:
\begin{equation}
     \partial^2_t \phi = \exp(2\kappa x)\partial^2_\perp \phi +\partial^2_x \phi.
\end{equation}
Assuming the plane wave form of
\begin{equation}
    \phi(x,y,z,t) = \hat{\phi}(x) \exp\left[i(k_y y+k_z z -\omega t) \right],
\end{equation}
it yields the Schr\"{o}dinger-type equation
\begin{equation}
[\omega^2-k^2_\perp \exp(2\kappa x)] \hat{\phi} +\partial^2_x \hat{\phi} =0,
\label{simplified_Schrodinger}    
\end{equation}
with $k^2_\perp \equiv k^2_y+k^2_z$. We note that in these Rindler coordinates that approximately describe the near-horizon geometry, the term $k^2_\perp \exp(2\kappa x)$ represents the angular momentum barrier of a black hole\footnote{To explicitly show this, as an example, let us consider a non-spinning black hole whose perturbation is governed by the Regge-Wheeler equation:
\begin{equation}
\left[ \frac{\partial^2}{\partial r^{\ast} {}^2} + \omega^2 -V (r) \right] \psi (r^{\ast}) = 0,
\end{equation}
where $r^{\ast}$ is the tortoise coordinate and $V(r)$ is the angular momentum barrier
\begin{equation}
V(r) \equiv \left( 1- \frac{2 M}{r} \right) \left( \frac{\ell (\ell+1)}{r^2} + \frac{2 M (1-s^2)}{r^3} \right).
\to k_{\perp}^2 e^{2 \kappa r^{\ast}} \ \ \text{for} \ r \simeq 2 M \ \text{and} \ l \gg 1,
\end{equation}
In the near-horizon limit, the potential barrier is approximated as $V(r) \to k_{\perp}^2 e^{2 \kappa r^{\ast}}$ with $k_{\perp}^2 \equiv \ell(\ell+1)/(2M)^2$, where we also assume $\ell \gg 1$ as we are interested in the contribution from higher harmonics. This is equivalent to the equation in (\ref{simplified_Schrodinger}). As long as we are interested in modes propagating near the horizon, in most cases (e.g. Kerr and Reissner-Nordstr\"{o}m black hole), perturbation equations with a short-range potential reduce to a simple form (\ref{simplified_Schrodinger}) regardless of the species of the field.}. Using the Wentzel–Kramers–Brillouin (WKB) approximation, we have
\begin{eqnarray}
    \hat{\phi}(x) \propto \frac{ \exp\left[i\int dx \cdot k_x(x)\right]}{\sqrt{k_x(x)}},\\
    k_x^2(x) \simeq \omega^2-k^2_\perp \exp(2\kappa x).
\end{eqnarray}
Setting $k^2_x > 0$ gives the maximum radius of the modes trapped by the angular momentum barrier:
\begin{equation}
    x<x_{\rm max} \equiv \kappa^{-1} \ln(\omega/k_\perp).  
\end{equation}

Let us next consider the position of a would-be horizon to evaluate $x_{\rm min}$. We here postulate that $x_{\rm min}$ is set by the interaction of the modes with the blue-shifted Hawking plasma near the horizon. To phenomenologically model this interaction, we shall add a dissipative term to the original wave equation \cite{Oshita:2018fqu,Oshita:2019sat}
\begin{equation}
    \partial^2_t \phi = (1+\gamma \partial_\tau)\left[\exp(2\kappa x)\partial^2_\perp \phi +\partial^2_x \phi \right],~~~ \partial_\tau \equiv \exp(-\kappa x) \partial_t, 
\end{equation}
where $\gamma$ controls the strength of the dissipation and $\partial_\tau$ is the derivative with respective to the proper time. In this model, the dissipation becomes significant when the frequency is blue-shifted and becomes $i\partial_\tau  = {\cal O}(1/\gamma)$. This yields the modified dispersion relation:
\begin{equation}
     k_x^2(x) \simeq \frac{\omega^2}{1-i\gamma \omega \exp(-\kappa x) }-k^2_\perp \exp(2\kappa x).
\end{equation}
Now near the horizon $x \to -\infty$, the transverse term (or angular momentum term) becomes negligible, but the dissipation term turns on and causes partial damping and reflection of the modes at 
\begin{equation}
    x \simeq x_{\rm min} = \kappa^{-1} \ln\left(\gamma \omega \right).
\end{equation}
Therefore, the WKB approximation should be valid in $x_{\rm min} <x < x_{\rm max}$, and thus the trapped modes in the cavity, whose length is $L_{\ell m} = x_{\rm max} - x_{\rm min}$, have normal modes with the following frequencies
\begin{equation}
    \omega_n \simeq k_x = \frac{n \pi}{x_{\rm max} - x_{\rm min}} =- \frac{ n \pi \kappa}{\ln(\gamma k_\perp)}. 
\end{equation}

Next, we shall assume that the gravitons with two polarizations are the most weakly coupled particles in nature, known as the weak gravity conjecture \cite{Arkani-Hamed:2006emk}, and thus gravitons survive as free particles even in the vicinity of the horizon. The entropy of a single graviton mode is then given by that of two bosonic harmonic oscillators in thermal equilibrium of Hawking temperature $T_{\rm H}$:
\begin{equation}
    S_n(k_y,k_z) =2 \times \left\{ \frac{\omega_n/T_{\rm H}}{\exp(\omega_n/T_{\rm H})-1}- \ln\left[ 1-\exp(-\omega_n/T_{\rm H})\right]  \right\}.
\end{equation}
The total entropy reads
\begin{align}
\begin{split}
    S_{\rm Q} &= 2 \times {\rm Area} ~\times  \int \frac{d^2k_\perp}{(2\pi)^2} \sum_n \left\{ \frac{ \omega_n/T_H}{\exp(\omega_n/T_H)-1}- \ln\left[ 1-\exp(-\omega_n/T_H)\right]  \right\},\\
    &= \frac{2\pi}{\gamma^2} \times {\rm Area} ~\times \int_0^\infty \frac{dw}{w^2} \left\{ \frac{ w}{\exp(w)-1}- \ln\left[ 1-\exp(-w)\right]  \right\} \sum^\infty_{n=0} n \exp\left[-(2\pi)^2 n \over w\right], \\
    &= \frac{\pi}{2\gamma^2} \times {\rm Area} ~\times \int_0^\infty \frac{dw}{w^2} \left\{ \frac{ w}{\exp(w)-1}- \ln\left[ 1-\exp(-w)\right]  \right\}\left[\sinh\left(2\pi^2/w\right)\right]^{-2}, \\
    &\simeq \frac{\rm Area}{55166\times  \gamma^2}.
    \end{split}
\end{align}
Equating this entropy with the Bekenstein-Hawking entropy of the horizon, we find the scale of the dissipation constant $\gamma$ as
\begin{equation}
    S_{\rm Q} = S_{\rm BH} = \frac{\rm Area}{4 G} \Rightarrow \gamma^2 \simeq \frac{4 G}{55166} \Rightarrow \gamma \simeq 8.52 \times 10^{-3} \times ({\rm Planck ~time}).
\end{equation}
Although the dissipation constant $\gamma$ is free to choose in our formalism, requiring the entropy $S_{\rm Q}$ is equivalent to the Bekenstein-Hawking entropy which says that the size of one bit is $\sim (\text{Planck length})^2$, it is found that the constant $\gamma^{-1}$ should be of the order of the Planck energy. We can also easily check that the dissipation and partial reflection happens at $x \simeq x_{\rm min}$, where the energy of gravitons reaches the maximum value of $\gamma^{-1} \sim (\text{Planck energy})$.

A partial reflection of propagating modes at $x=x_{\rm min}$ leads to the emission of gravitational wave echoes as shown in the literature \cite{Abedi:2016hgu,Abedi:2018npz,Oshita:2018fqu,Oshita:2019sat,Wang:2019rcf,Abedi:2020ujo}. The interval time of the signals is given by the echo time $t_{\rm echo}$, which can be expressed in terms of the entropy of the quantum black hole $S_{\rm Q}$ as
\begin{align}
\begin{split}
t_{\rm echo} &= 2(x_{\rm max} -x_{\rm min}) = -\kappa^{-1} \ln(\gamma^2 k_\perp^2),\\ 
&\simeq \frac{1}{2\pi T_H}\ln\left[ {\rm Area} \over 4\pi \ell(\ell+1) \gamma^2 \right] \simeq \frac{1}{2\pi T_H}\ln\left[55166 S_{\rm Q} \over 4\pi \ell(\ell+1)\right],
\end{split}
\label{echo_time}
\end{align}
where we used $\pi k^2_\perp \times {\rm Area} \simeq (2\pi)^2\ell (\ell+1)$ for an approximate correspondence between the number of spherical and flat harmonic modes. The formula of the echo time is similar to that of the scrambling time \cite{Sekino:2008he} that quantifies how fast the information in a black hole is mixed up. Our computation predicting the position of the dissipative region, $x=x_{\rm min}$, is, in this sense, consistent with the black hole information recovery scenario \cite{Sekino:2008he}.

\section{Discussions and conclusions}
\label{sec_discussions}
We estimated black hole entropy by assuming that it is in quasi-thermal equilibrium with the Hawking temperature and that each quasi-normal (QN) mode contributes to the partition function with the weight of the Boltzmann factor.
We found that the entropy estimated from the QN modes for a classical black hole, whose horizon is a perfect absorber, is of the order of unity. On the other hand, a quantum black hole which has the reflective boundary due to the viscosity near the horizon leads to entropy comparable to the Bekenstein-Hawking formula.
The frequency of a QN mode for each angular mode, for large $\ell$, is given by
\begin{align}
\omega_{\ell mn} \simeq \frac{8 \pi T_{\rm H}}{3 \sqrt{3}}  \times
\begin{cases}
\ell -i (n+1/2) \ &\text{for} \ \text{a classical black hole,},\\
\displaystyle \frac{3 \sqrt{3} (n+1)}{8 \pi \ln{(S_{\rm Q}/\ell^2)}} \ &\text{for} \ \text{a quantum black hole},
\end{cases}
\end{align}
where $(\ell,m)$ is the label of angular mode, $S_{\rm Q}$ is the thermal entropy of the modes. To find a finite entropy for quantum black holes,  we here introduced the energy scale $1/\gamma$, above which the modes are damped significantly. Computing the partition function of QN modes with the weight of $e^{-\tilde{\omega}_{\ell mn}/T_{\rm H}}$ with $\tilde{\omega}_{\ell mn} \equiv |\omega_{\ell mn}|$, we find $S_{\rm classical} \sim {\cal O} (1)$ and $S_{\rm quantum} \sim {\cal A}/G$,  if we set the damping scale to the Planck energy $1/\gamma \sim M_{\rm Pl}$ . This is because for a classical black hole, only a few harmonic modes contribute to the partition function while many modes do so for a quantum black hole (see Figure \ref{fig:mode_count}). This is because $\tilde{\omega}_{\ell mn}/T_{\rm H} \sim \ell$ for a classical black hole, which rises much faster than $\tilde{\omega}_{\ell mn}/T_{\rm H} \sim 1/ \ln(S_{\rm BH}/\ell^2)$ for a quantum black hole. In the latter case, the Boltzmann factor does not suppress the excitation of QN modes up to $\ell^2 \lesssim S_{\rm BH}$ unlike the former case. Note that our way of modelling a quantum black hole is not unique, as dissipation is a simple phenomenological toy model for interaction with BH microstates. An important lesson is that quantum effects near the horizon may significantly change the configuration of QN modes as was reported in echo model (see also Ref. \cite{Jaramillo:2020tuu} for the pseudospectrum of QN modes), and higher angular modes play a key role in reproducing to the Bekenstein-Hawking area law of entropy.

An interesting corollary of our finding is the possibility of simulating horizon entropy and echoes within analog black holes in condensed matter systems, such as the $8Pmmn$ borophene sheet  \cite{Farajollahpour:2019kwj}. Although there are other material that may mimic a horizon, e.g., liquid Helium \cite{Novello:2002qg} or Bose-Einstein condensates \cite{Macher:2009nz,Curtis:2018qey}, they require very low temperature or high pressure. More importantly, they involve a dynamical flow and thus are not near thermal equilibrium. On the other hand, the $8Pmmn$ borophene sheet can provide an analog horizon at ambient, near-equilibrium, conditions. Applying perpendicular electric field to a portion of the sheet, the region with the strong field has a strongly tilted Dirac cone, which corresponds to a boosted light cone in a gravitational system. Overtilted Dirac cones can mimic the interior horizon or the Killing horizon of a black hole. Our findings suggest that by including the dissipation effects in the system, it could be possible to demonstrate analogues of horizon entropy and echoes on the tabletop. In the two-dimensional black hole, the angular momentum potential does not exist. As such, we may need to add a potential barrier by applying inhomogeneous external electric field to reproduce a potential outside the analog horizon. Then the resonance in the cavity between the potential and the analog horizon may lead to multiple QN modes, which comprise the entropy of the horizon, according to our calculations. One could thus measure the echo time and entropy of the system to test the relation in (\ref{echo_time}).

\begin{acknowledgements}
We are thankful for valuable discussions with Seyed Akbar Jafari, Matt Johnson, and Jiayue Yang. This research was funded thanks in part to the Special Postdoctoral Researcher (SPDR) Program at RIKEN (NO), FY2021 Incentive Research Project at RIKEN (NO), and Grant-in-Aid for Scientific Research (KAKENHI) project for FY 2021 (21K20371) (NO), the Natural Sciences and Engineering Research Council of Canada, and the Perimeter Institute (NA). Research at Perimeter Institute is supported in part by the Government of Canada through the Department of Innovation, Science and Economic Development and by the Province of Ontario through the Ministry of Colleges and Universities.
\end{acknowledgements}

\end{document}